\documentclass[11pt]{article}
\usepackage{url}
\usepackage{epsfig}

\begin{document}

\centerline{\Large \bf Towards the  quantification of the semantic information}
\centerline{\Large\bf  encoded in written language}

\vskip 1cm
\centerline{ Marcelo A. Montemurro}
{ \it
\centerline{Faculty of Life Sciences, The University of Manchester, M13 9PT, Manchester,}
\centerline{United Kingdom}
\centerline{m.montemurro@manchester.ac.uk}} 
\vskip 0.5cm
\centerline{Dami\'an Zanette}
{\it
\centerline{Consejo Nacional de Investigaciones Cient\'{\i}ficas y T\'ecnicas }
\centerline{ Centro At\'omico Bariloche and Instituto Balseiro, 8400 San Carlos de Bariloche, R\'{\i}o Negro,  Argentina}
\centerline{zanette@cab.cnea.gov.ar}}

\vskip 1cm

\begin{abstract}
Written language is a complex communication signal capable of
conveying information encoded in the form of ordered sequences of
words. Beyond the local order ruled by grammar, semantic and
thematic structures affect long-range patterns in word usage. Here,
we show that a direct application of information theory quantifies
the relationship between the statistical distribution of words and
the semantic content of the text. We show that there is a
characteristic scale, roughly around a few thousand words, which
establishes the typical size of the most informative segments in
written language. Moreover, we find that the words whose
contributions to the overall information is larger, are the ones
more closely associated with the main subjects and topics of the
text.  This scenario can be explained by a model of word usage that
assumes that words are distributed along the text in domains of a
characteristic size where their frequency is higher than elsewhere.
Our conclusions are based on the analysis of a large database of
written language, diverse in subjects and styles, and thus are
likely to be applicable to general language sequences encoding
complex information.
\end{abstract}


\section{Introduction}

Among the higher functions of our brain, language stands as a unique
ability conferring us a distinctive trait from the rest of living
beings \cite{deacon1997,smith1999}.  Language has developed
following the laws of natural evolution
\cite{lieberman2007,nowak2002,nowak2000} with the functional goal of
encoding and transmitting information between humans. Although the
information to be encoded is usually highly complex, it can be
readily projected onto a string of words. It has been argued that
this is possible due to the presence of long-range memory in word
sequences, which in turn creates organisational structures that go
beyond the scope of sentences and paragraphs, and can extend for
hundreds or thousands of words \cite{lacalle2006}.  Although, the
suggestion of statistical macro-structures associated with the
semantic content of linguistic communication has been suggested
before \cite{kintsch1978}, it has not yet been accounted for with
the use of information theory.

In recent years the use of tools drawn from statistical physics has
quantitatively revealed rich linguistic structures at many scales,
ranging from the domain of syntax to the organisation of whole
lexicons and literary corpora
\cite{cancho2001b,cancho2001,cancho2003,dorogovtsev2001,ebeling1994,montemurro2002,sigman2002,zanette_Montemurro2005}.
However, a fundamental question that has not directly been addressed
so far is how statistical structures relate to the function of
encoding complex information.  The earliest attempt to address this
question was probably due to Claude Shannon, soon after the
inception of information theory \cite{shannon1948}. He represented
language as a communication channel encoding information in the
sequence of its basic symbols, which in his analysis were 26 letters
and the blank space, and obtained an estimate of the entropy of
written English \cite{shannon1951}. However, his analysis was not
designed to relate an information measure to the semantic function
of language. Instead, it was aimed at reflecting the statistical
structure of linguistic sequences independently of the specific
information that was being encoded.  More recently, we showed that
an entropy measure of the word distribution over a text bears
information about the specific linguistic role of some word classes
\cite{montemurro_zanette2002}. This approach disclosed that, for
instance, adverbs and most verbs tended to be more uniformly
distributed than nouns or pronouns. Thus, simply by associating the
entropy measure with a word, it was possible to know whether that
word was, for example, more likely an adverb or a noun.

Here, we take a significant step further and propose a measure,
based on Shannon's mutual information \cite{shannon1948}, that
captures the relationship between the statistical structure of word
sequences and their semantic content. First, we show that words
typically appear distributed in domains, so that in certain sections
of the text the frequency of a word is consistently higher than
elsewhere. This structured heterogeneity in the distribution of
words encodes information about the sections of the text in which
the words appear. We then use information theory to quantify the
relationship between the distribution of words and the sections in a
given partition of the text. We find that there is a typical size of
the parts for which the information in word distribution is maximal,
thus revealing a characteristic scale that can be shown to be
related to the semantic content of language. This scale is built up
from contributions of all the different individual words, as a
direct consequence of the domain structure in their distributions.
Finally, we show that the words that contribute the most to the
total information are those more closely related to the main topics
and subjects if the text.

\section{Domain structure in word distribution}
Although it is evident that different words appear with variable
frequency in different parts or sections of a text, it has only
recently been pointed out that the patterns of frequency variability
are related to the linguistic role of words
\cite{montemurro_zanette2002}. Additionally, by analyzing the
statistics of the distance between consecutive occurrences of the
same word in a text, it was shown that many words exhibit a
phenomenon of clustering or ``self-attraction''
\cite{herrera2008,ortuno2002}.  Below, we quantify these results
within the framework of information theory and show that the
long-range distribution of words in written language bears the
imprint of the semantic content encoded in the text. First, however,
we show that the variability in the frequency of word usage is
characterized, for many words, by the presence of a series of
domains where the frequency of use of the word is higher than
outside the domains. Moreover, these domains have a typical size
that depends on the specific word.  As we shall see below, they
contribute to determining the scale of the most informative
structures in written language.

In a given text of $N$ words in length, we denote the position of
every word by the variable $t = 1,\ldots, N$. Then, we can represent
the occurrences of a particular word $w$ by a sequence of sharply
peaked functions located at the positions where the word $w$
appears.  A local average of this sequence, defined as a convolution
with a narrow bell-shaped function, allows us to define a rate of
occurrence of word $w$  (see Appendix A).

\begin{figure}
\centerline{\psfig{file=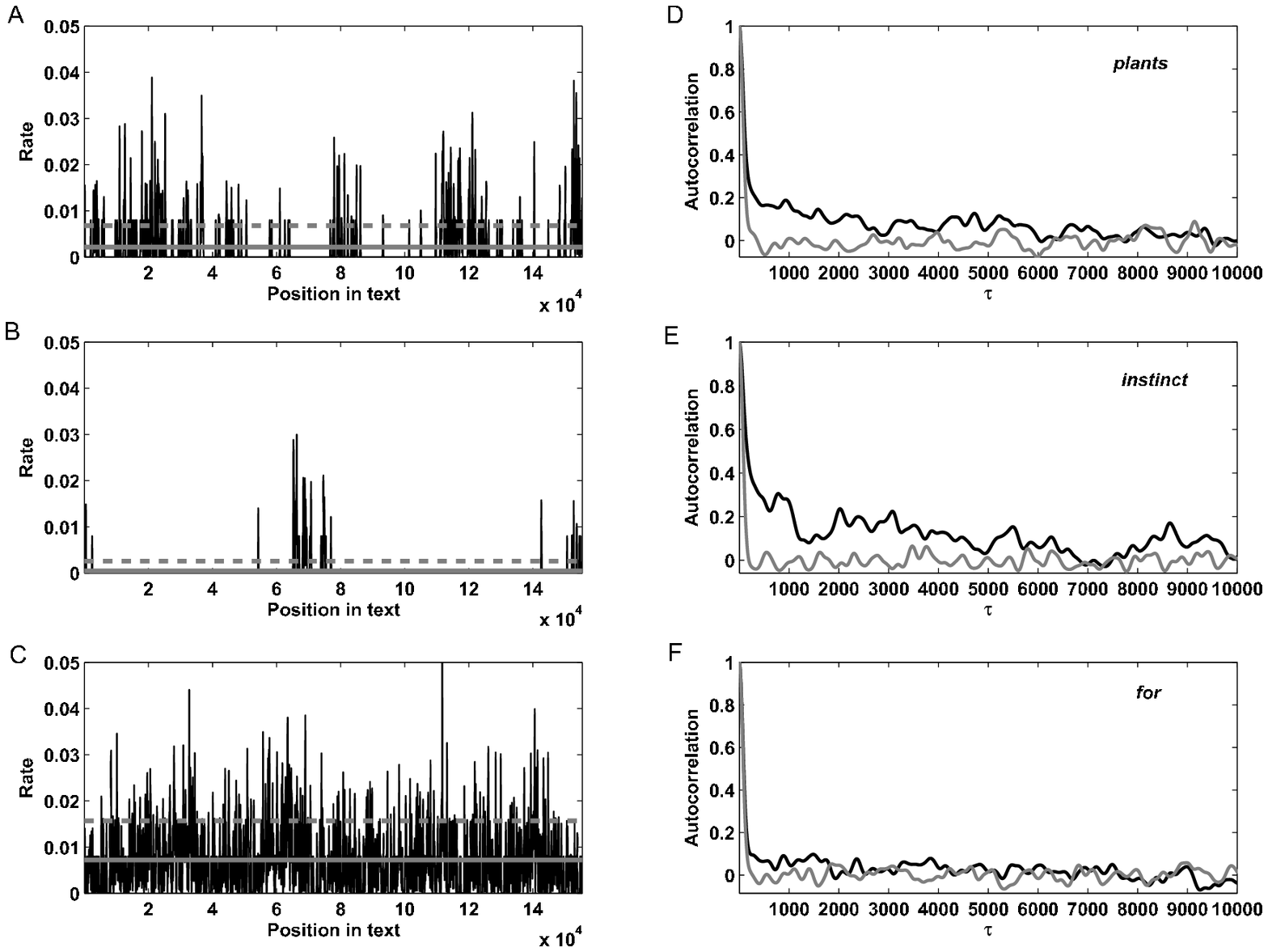,width=15cm}} \vspace*{8pt}
 \caption{{\bf Distribution of three words in  {\it The
Origin of Species}, by Charles Darwin.} Left panels (A, B, and C)
show the rate of occurrence of the words {\it plants}, {\it
instinct}, and {\it for} as a function of the position in the text
measured in number of words (black line), its average (gray full
line) and one standard deviation above the average (grey dash line).
Right panels (D, E, and F) show the autocorrelation of the rate of
occurrence of the same words as a function of the distance between
text positions. The black line shows the autocorrelation for the
word as it appears in the original text. The gray line corresponds
to a random shuffling of all words in the text.}
\label{fig1}
\end{figure}

The left panels in Fig. \ref{fig1} show the rate of occurrence of
three words from {\it The Origin of the Species}, by Charles Darwin.
Panel A shows  for the word {\it plants}. With its 335 occurrences,
this word is considerably frequent in {\it The Origin of Species},
but its use is strongly heterogeneous. It tends to appear in
localized sections of the text, exhibiting higher frequencies within
a number of domains spread throughout the text. The second word we
analysed is {\it instinct}, shown in panel B. This word appears 69
times, with strongly localized occurrences limited to essentially
two sections of the text. The region where it occurs with the
highest frequency spans approximately 104 words. The third word,
shown in panel C, is {\it for}. This is a common English word,
appearing 1123 times in the text. Its usage is not directly linked
to any specific thematic context. Therefore, apart from seemingly
random fluctuations, the rate at which for is used is roughly
uniform.

The presence of domains in the distribution of words can be
quantitatively revealed by computing the normalized autocorrelation
of the rate of occurrence, defined as
\begin{equation}
c_w(\tau)=\frac{\langle\rho_w(t)\rho_w(t+\tau)\rangle_t-\langle\rho_w(t)\rangle_t\langle\rho_w(t+\tau)\rangle_t}{\langle\rho_w(t)^2\rangle_t-\langle\rho_w(t)\rangle_t^2}
\end{equation}
where $\langle \ldots \rangle_t$ indicates an average taken over all
text positions. If the occurrence of a particular word $w$ is
concentrated in localized domains, the autocorrelation function will
be significantly different from zero up to a value of $\tau$ of the
order of the typical domain size \cite{bray1994}. Panels D, E, and F
show the autocorrelation computed for the words of the left-hand
panels (black curve). We also show the autocorrelation obtained for
the same words after all the words in the text have been randomly
shuffled, thus destroying any pattern in word appearance (gray
curve).

The autocorrelation for the word {\it plants }(panel D) shows that its
usage is organized into spans extending 2000-3000 words, in
agreement with the domain structure shown in panel A. A similar
situation occurs in the case of {\it instinct}, for which correlation
structures extend over a scale of approximately 5000 words. On the
other hand, the autocorrelation for the word {\it for} falls to the random
level at very short distances, thus indicating no pattern of domains
in its usage.

Heterogeneity in the rate of word occurrence over the text is
directly related to the specificity of that word to certain sections
of the text. For instance, the use of the word {\it instinct} directly
tags a few parts where, due to the specific subject being treated
there, it occurs more frequently than elsewhere in the text. If the
word {\it instinct} is found in the text, it is likely that that
particular occurrence belongs to one of the sections in which the
rate of instinct is higher.  Likewise, the distribution of {\it plants}
also offers hints about which part of the text the word is found in.
The variability in the frequency of word usage can therefore be
exploited to discriminate between different sections of the text.
Our aim in the following is to quantify how efficiently this
discrimination can be carried out, giving a statistical measure of
the amount of knowledge that the word distribution carries about the
parts of the text where each word appears.

\section{Information in the distribution of words}
As discussed above, the heterogeneity in the distribution of
individual words can be used to tag different sections of the text.
In what follows, we consider a text divided into contiguous sections
of equal length, and then define an information-theoretic measure
that captures the information contributed by the distribution of all
individual words in the text. We then show that this information is
maximal for a characteristic size of the parts into which the text
is divided --typically, of a few thousand words. We thus show that
information theory can be used to characterize a new scale for
linguistic structures, much longer than the range of application of
grammar rules but still much shorter than, for instance, most books.
As we shall see below, this scale turns out to be related to the
size of semantic domains in written language.

Consider a text of $N$ words in length, and with a lexicon of $K$
different words. In our analysis, two words are considered to be
different if they are spelt differently, in particular, even if they
are inflected variations sharing a common root. We divide the text
into $P$ parts of identical length, each part containing $s = N/P$
words. Our aim here is to define an information-theoretical measure
to quantify the relationship between the heterogeneities in the
distribution of words due to their linguistic function and the text
partition.

We define our measure of relative information as the difference
between Shannon's mutual information \cite{cover2006,shannon1948}
evaluated in a given text and in a surrogate version of it. The
surrogate text is built form the original one by randomly shuffling
all its words. In this way, the random text will have exactly the
same overall word frequencies as the original text, but will lack
any linguistically relevant order in the sequence of words. The
basic assumption behind our definition is that the constraint of
encoding linguistic information determines the degree of order by
which the original text exceeds its random counterpart. In other
words, our measure establishes the amount of information required to
re-arrange the words of the random text to recover the word
distribution over the parts of the original text. The more
heterogeneous the word distribution in the original text, the more
information will be required to re-order its words from the random
shuffling. A direct application of Shannon's mutual information
leads to the following definition (see Appendix B):
\begin{equation}
\label{DI}
\Delta I(s)=\sum_{w=1}^{K} p(w) \left[ \langle \hat{H}(J|w)
\rangle-H(J|w) \right]  ,
\end{equation}
where the sum runs over the lexicon of the text in question, i.e.
over its $K$ different words. For a word $w$ with a total of $n$
occurrences, the coefficient $p(w) = n/N$ gives its frequency over
the whole text. The entropy-like quantity $H(J|w)$  is given by
\begin{equation}
\label{entropy} H(J|w)=-\sum_{j=1}^P\frac{n_j}{n}\log_2 \frac{n_j}{n}
,
\end{equation}
where $n_j$ is the number of occurrences of word $w$ in part $j$ of
the real text, and the sum runs over all the parts. Meanwhile,
$\langle \hat{H}(J|w)\rangle$ is the same quantity calculated for a
random shuffling of all the words in the text, and averaged over the
all possible realizations of the random permutation of words. This
average can be computed analytically (see Appendix C).

\begin{figure}
\centerline{\psfig{file=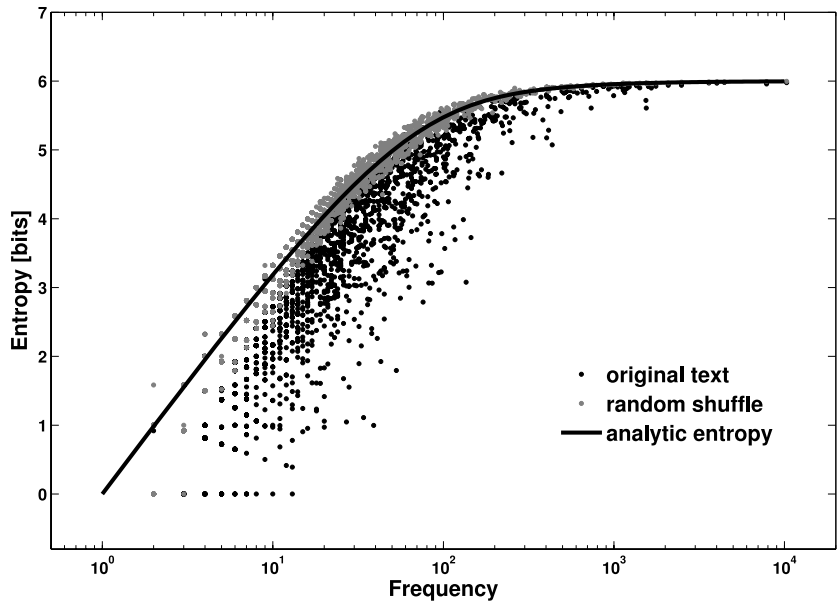,width=11cm}} \vspace*{8pt}
 \caption{{\bf Word entropy for a real and a random
text}. The black dots show a scatter plot of the entropy in bits for
all the words in {\it The Origin of the Species} as given by Eq.
(\ref{entropy}) and using a partition of the text in $P=64$ parts.
The gray dots show the entropy of the words in randomly shuffled
realization of the original text. The full line is the result of the
analytical expression for the entropy (see Appendix C) of the words
in the random version of {\it The Origin of Species}.}
\label{fig2}
\end{figure}

The behavior of the entropy quantities can be illustrated with an
example. In Fig. \ref{fig2} we show the value of the entropy for all
the words appearing in { \it The Origin of Species} as a function of
word frequency (black dots). We also plotted the entropy for the
words in one realization of the random permutation of word positions
(grey dots), and the value of the same entropy when it is averaged
over an infinite number of realizations of the random text (black
line). The latter corresponds to the exact analytical calculation
of the entropy.

There are two important observations in regard to the comparison
between the entropy of the real text and that in the random version.
First, the entropy of words in the real text is on average lower
than in the random version of the text for the same frequency range.
The lower value of the entropy for words in the real text is a
consequence of the more structured distribution of words sequences
conveying complex information in contrast to the randomly located
words in the shuffled text. Second, the fluctuations across
different words in the same frequency range vary over a much larger
range in the real text than in the shuffled one. This is due to the
fact that the large variations in word distribution are not
stochastic but imposed by linguistic and thematic constraints in the
text. Therefore, even groups of words having the same frequency can
show sharp variations in their patterns of occurrence due to their
different linguistic roles.

For a specific text, the relative information defined by Eq.~ \ref{DI}
depends only on $s$, the number of words per part, through the
number of parts $P$.  The variable $s$ sets a scale for the
coarseness with which the distribution of words over the text is
determined. At a given scale, $\Delta I(s)$ quantifies how much
information about specific parts of the text is contained in the
distribution of words with respect to that in the random text. Words
that tend to appear evenly over the text will contribute little
information about different parts. On the other hand, words with
non-uniform distributions that cannot be simply associated with
random fluctuations will have a significant contribution to the
total relative information. As given by Eq.~\ref{DI}, $\Delta I(s)$ will have units
of bits per word.

\begin{figure}
\centerline{\psfig{file=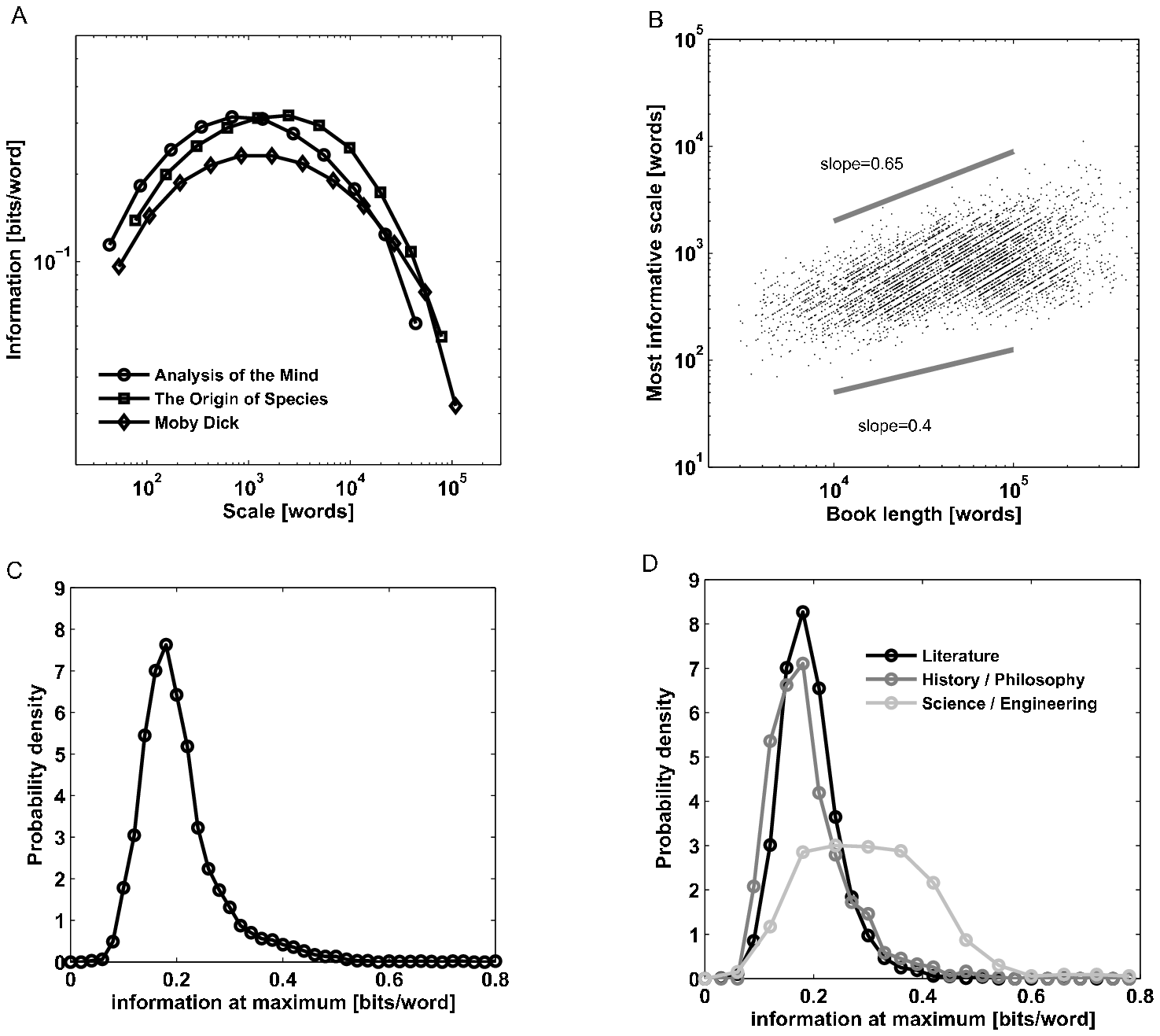,width=15cm}} \vspace*{8pt}
 \caption{{\bf Information encoded in whole texts.} A.
Information $\Delta I(s)$ in bits per word for three texts, as a
function of the scale $s$. The texts are {\it Analysis of the Mind},
by Bertrand Russell (circles); {\it The Origin of Species}, by
Charles Darwin (squares); and {\it Moby Dick}, by Herman Melville
(diamonds). B. Scatter plot of scale at which the total information
is maximal as a function of the text length for all the texts in a
corpus of 5258 books in English. The straight lines are power-law
functions (note the logarithmic scales) whose slopes approximately
confine the speed of growth of the most informative scale with text
length. C. Normalized histogram showing the distribution of the
maximum total information for all the books in the corpus. D.
Normalized histogram showing the distribution of the maximum total
information for each of the three partitions in which the corpus was
divided. The number of books in each division of the corpus was as
follows: literature, 3329 books; history and philosophy, 1374 books;
science and engineering, 555 books. }
\label{fig3}
\end{figure}

As an example, we describe  as a function of the scale $s$ for three
texts: {\it The Origin of Species} by Charles Darwin, {\it Analysis
of the Mind} by Bertrand Russell, and {\it Moby Dick} by Herman
Melville.  The lengths of the texts were, respectively, 155800,
89586, and 218284 words. Figure \ref{fig3}A shows that, in all three
texts, there is a scale at which the information $\Delta I(s)$ is
maximal. In these examples, the maximal information values occur at
a scale of approximately 3000 words for {\it The Origin of the
Species}, 700 words for {\it Analysis of the Mind} and 1200 words
for {\it Moby Dick}. For each text, starting from scales of around
100 words, the information encoded in the word distribution
increases with s up to a specific most informative scale. This means
that the distribution of words better discriminates between
different sections of the text for a characteristic size of the
parts. It is for this particular size that the mutual information of
the word distribution and the text partition differs maximally
between the real and the random texts. The scales at which the
information is maximal, roughly between 1000 and 3000 words in
length, are much larger than the scope of grammatical rules. As we
shall argue below, they are rather related to the semantic structure
of the texts.

For the three texts considered above, there is no direct relation
between text length and the size of the most informative scale.
However, in principle, longer texts allow their semantic structures
to span longer scales. To investigate this hypothesis more in
detail, we analysed a corpus consisting of 5258 books in English,
including English and American literature, as well as texts on
science, engineering, technology, history, philosophy, and religion.
All the texts were obtained from the Project Gutenberg internet site
\footnote{www.gutenberg.org}. In Fig. \ref{fig3}B we show a scatter
plot of the scale at which the information $\Delta I(s)$ is maximal
versus the length of the text, for all of the books in our corpus.
Although the data exhibit a large dispersion, there is a clear
pattern of slow growth of the most informative scale with text
length. As a guide to the eye, we also plotted two straight lines
with slopes 0.65 and 0.4 which, in this log-log plot, correspond to
power-law functions. The most informative scale of books around 105
words in length typically lies between 300 and 3000 words. This
means that, for such texts, the size of the parts at which the
information $\Delta I(s)$ attains its maximum is roughly 100 times
less than the text length.

It is interesting to identify which are the books that fall in
extreme regions of the cloud of points in Fig. \ref{fig3}B. For
instance, at the lower-left corner, which corresponds to short books
with a small most informative scale $s$, we find works like {\it
Quotations of Lord Chesterfield}, {\it Quotes and Images From The
Novels of Georg Ebers}, which consist of a collection of short
quotations with no thematic unity building up along the text.
Consistently, the scale associated to these books lies in the range
between 50 and 70 words. At the opposite extreme, at the upper-right
corner of the cloud of points, which corresponds to long books and
large most informative scales, we find long treatises on subjects
with clear thematic unity. In particular, the three works with the
largest most informative scale were {\it History of The Decline and
Fall of the Roman Empire}, Vol. 3, by E. Gibbon, {\it A History of
Rome}, Vol. 1, by A. Greenidge, and {\it Civilization of the
Renaissance in Italy}, by J. Burckhardt.

Notwithstanding the large dispersion of the most informative scale,
the maximum value of the information $\Delta I(s)$ for each book was
surprisingly consistent across the corpus.  Figure \ref{fig3}C shows
a histogram of the maximum information for the books in the corpus.
It turns out to be narrowly centred at 0.2 bits/word. This suggests
that, in written English, words tag different parts of the text with
very consistent accuracy, although the size of those parts may vary
substantially between texts.  However, when looking at the books
within a given range of values of the maximum information per word
we found a tendency for literary and history books to have lower
values of the information per word than many of the books on science
and engineering. The pattern becomes evident in Fig. \ref{fig3}D
where we show the normalized histograms of the maximum information
computed on three subsets of the whole corpus. Whereas the
information for the literary, and history and philosophy books has a
very similar distribution characterized by a narrow peak around 0.2
bits/word, the values for science and engineering books have a
broader distribution that extends over higher values of the
information. This shows that $\Delta I(s)$ can quantify the
similarities and difference in language styles in the different
components of the corpus. In particular, it indicates that the usage
of language in scientific and engineering books is such that the
distribution of words tags more efficiently the different parts of
the text.

\section{The information of individual words}
We deal, in this section, with the contribution of individual words
to the information $\Delta I(s)$. It turns out that each word has
its own characteristic scale, at which the information it provides
about specific parts of the text is maximal. By means of a heuristic
model, we show that the existence of that scale is a consequence of
the domain structure of word distribution.

Equation~\ref{DI} shows that $\Delta I(s)$ is additive over the whole
lexicon. Thus, each term in the sum can be interpreted as the
contribution of a single word to the total information. For a word
$w$, its contribution to the total information equals $\Delta
I_w(s)=p(w)\left[\langle\hat{H}(J|w)\rangle-H(J|w)\right]$. This
means that we can not only compute the overall information as a
function of the scale $s$, but also associate a measure of
information with every different word appearing in the text. Thus,
by using the single-word information $\Delta I_w(s)$, it is possible
to identify the most informative words in a given text. Remarkably,
as we shall illustrate below for {\it The Origin of Species}, the most
informative words coincide with those words that any human reader
would choose as most representative of the subject of the text.
Therefore, summing the contributions of every word, the total
information $\Delta I(s)$ can be interpreted as a measure of the
overall semantic information in the text in question.

The mathematical form of $\Delta I_w(s)$ also suggests another
interpretation of the meaning of the overall information $\Delta
I(s)$. The contribution of each word to the total information is
proportional to both the frequency of the word and to the difference
in entropy of the word distributions in the random and real texts.
The difference in entropies is a direct measure of the degree of
order that exists in the word distribution in the real text beyond
that determined by the word frequencies alone. Ultimately, that
order has an origin in the semantic role of the word.  That means
that for a word of a given frequency $p(w)$, the more heterogeneous
the distribution of the word compared to its stochastic counterpart
in the random text, the larger its contribution to the total
information.

\begin{figure}
\centerline{\psfig{file=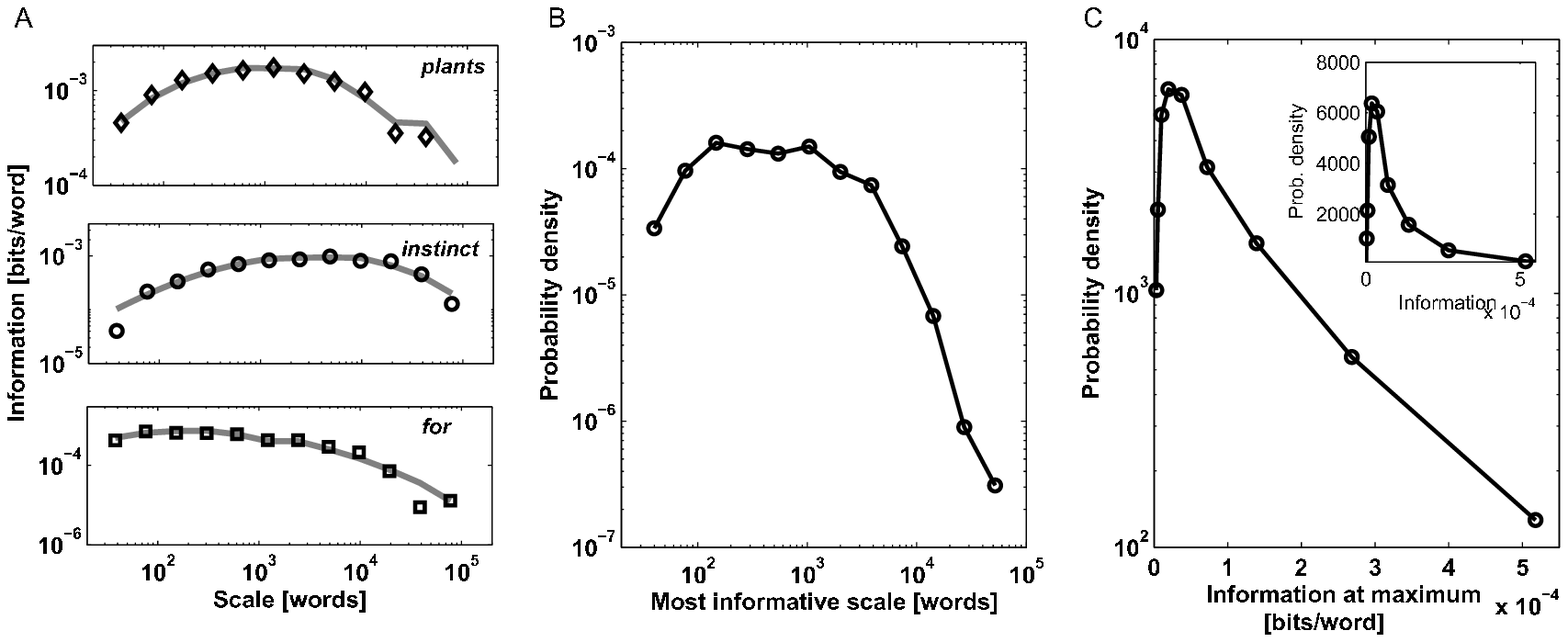,width=15cm}} \vspace*{8pt}
 \caption{{\bf Information encoded in single words.} A.
The three subpanels show the information of single words for the
three examples shown in Fig. \ref{fig1}, as a function of the scale
$s$. The full lines are fittings obtained using the model described
in the text. B. Probability density of the most informative scale
for all the words in the The Origin of the Species. C. Probability
density of the maximum information for all the words in {\it The
Origin of the Species}. }
\label{fig4}
\end{figure}

To capture the role of individual words in determining the semantic
information of the text, we analysed the information encoded in
single words as a function of the scale $s$. In Fig. \ref{fig4} we
show the results obtained for {\it The Origin of Species}.

Symbols in Fig. \ref{fig4}A stand for the single-word information
$\Delta I_w(s)$ as a function of $s$ for the words {\it plants,} {\it instinct},
and for. The profile of $\Delta I_w(s)$ vs. $s$ is qualitatively the
same for the three words, with a maximum at an intermediate scale.
The maximum is located at 1216 for {\it plants}, at 4864 for {\it instinct},  and
at 76 for the word {\it for}. Thus, the most informative scale is
different for each word, but roughly lies in the interval where the
total information $\Delta I(s)$ of different texts attains its
maximum (see panels A and B in Fig. \ref{fig3}). We show below that
this scale can be related to the size of typical domains in the
distribution of each word. The most informative scale of a whole
text is ultimately determined by the interplay between the positions
of the maxima of its individual words.

In Fig. \ref{fig4}B we have plotted a normalized histogram of the
most informative scale for all the words in the book. It exhibits a
flat maximum spanning scales between 100 and 1000 words
approximately, after which it decays sharply.  In Fig. \ref{fig4}C
we show a histogram of the value of the maximum information for the
individual words. As it was the case in Fig. \ref{fig3}C, the
distribution is narrowly cantered around a typical information value
of about bits/word.

The behavior of the single-word information as a function of the
scale $s$ can be heuristically explained by means of a simple model.
It consists of a stochastic representation of the distribution of
the $n$ occurrences of a given word over the text of length $N$. Our
main assumption is that the word in question has a uniform random
distribution all over the text, except for a localized domain where
its frequency is larger than in the rest. The distribution is
specified by giving the length $L$ of the domain where the word is
more concentrated, the place $N_0$ in the text at which the domain
begins, and the excess ne of occurrences of the word in the domain.

Once the three parameters $L$, $N_0$, and $n_e$ have been specified,
the distribution is built by allocating the $n$ occurrences of the
word along the text, following two steps: (i) A number $n-n_e$ of
occurrences are uniformly distributed at random all over the text.
(ii) The remaining ne occurrences are uniformly distributed at
random over the concentration domain, i.e. between places $N_0$ and
$N_0+L-1$. Of course, only one occurrence per place is allowed.
Also, the domain length $L$ is supposed to be large enough as to
allow for the ne excess occurrences once step (i) is completed. The
expected number of occurrences inside the domain is $n_e+(n-ne)L/N$
while, in the remaining of the text, any portion of length $L$
contains, on the average, just $(n-n_e)L/N$ occurrences. The random
distribution is build numerically, and the value of $\Delta I_w(s)$
as a function of the number of words per part is averaged over
realizations of the allocation process.

The effect of varying any of the three parameters $L$, $N_0$, and
$n_e$, leaving the other two fixed, can be qualitatively described
as follows. The domain length $L$ controls the position of the
maximum of $\Delta I_w(s)$, which roughly coincides with $L$.
Simultaneously, since as $L$ grows the difference in frequency
between the domain and the remaining of the text decreases, larger
values of $L$ imply smaller values of $\Delta I_w(s)$. Conversely, a
growth in the excess ne increases the difference in frequency, and
therefore implies a growth in $\Delta I_w(s)$. Finally, changes of
the domain position upon variation of $N_0$ have practically no
effect on the position and height of the maximum of $\Delta I_w(s)$,
but control the form in which $\Delta I_w(s)$ decays at both sides
of the maximum.

The grey curves in the three sub-panels of Fig. \ref{fig4}A show the
single-word information for {\it plants}, {\it instinct}, and {\it
for}, estimated using our model. The agreement is excellent,
strongly suggesting that the domain-like distribution pattern of
words along the text is the most relevant factor determining the
characteristic behavior of $\Delta I_w(s)$.


\begin{table} \label{tab1}

\caption{Most informative words for three books.}

\center{\begin{tabular}{@{}|c|c|c|@{}} 
\hline
The Origin of Species &
Analysis of the Mind & Moby Dick\\ 
\hline
on &image&I\\
species&memory&whale\\
varieties&images&you\\
hybrids&word&Ahab\\
forms&belief&is\\
islands&words&ye\\
of&desire&Queequeg\\
will&sensations&thou\\
selection&object&me\\
genera&you&of\\
plants&past&he\\
seeds&knowledge&captain\\
sterility&box&boat\\
fertility&content&the\\
characters&consciousness&Stubb\\
breeds&appearances&his\\
groups&movements&Jonah\\
water&mnemic&was\\
the&feeling&whales\\
formations&proposition&my\\
pollen&general&him\\
bees&particulars&Starbuck\\
instincts&thought&sir\\
new&experience&white\\
he&objective&sperm\\
rudimentary&meaning&bildad\\
cells&laws&her\\
organs&introspection&we\\
intermediate&animal&Peleg\\
crossed&vague&said\\
natural&sensation&fish\\
birds&physical&Pip\\
would&habit&old\\
I&the&a\\
domestic&matter&cook\\
wax&we&flask\\
formation&response&aye\\
organ&propositions&ship\\
\hline

\end{tabular}}
\end{table}

Finally, we provide evidence that the information captured by
$\Delta I_w(s)$ is related to the semantic role of words. With this
aim, we computed the single-word information for all the words in
{\it The Origin of the Species}, in {\it Analysis of the Mind}, and
in {\it Moby Dick}, at the scale $s$ where the total information
$\Delta I(s)$ was a maximum for each of the texts (see Fig.
\ref{fig3}A). In Table 1, we present the first words of
the three texts ranked by the information $\Delta I_w(s)$ encoded by
each word.  Note, from Eq.ref{DI}, that the contribution of each
word to the total information is weighted by its frequency in the
whole text, $p(w)$. Nonetheless, very few of the most frequent words
--such as {\it the}, {\it and}, {\it of}, {\it or}-- appear at the
top of the lists. This is because, in real texts, the most common
words have distributions which do not differ much from that of a
typical realization of the randomly shuffled text.  Remarkably, on
the other hand, the words with largest information are specifically
relevant to the main subjects of the text. Among the top ten words
of {\it The Origin of Species}, for instance, we find {\it species},
{\it varieties}, {\it hybrids}, {\it forms}, {\it islands}, {\it selection} and {\it genera}. Those
words would certainly be signalled out by a reader as some of the
most representative of the message conveyed by the text.   In {\it
Analysis of the Mind} we recognize a very similar situation. Its top
words are essential to the philosophical subject of the book. The
comparison with {\it Moby Dick} is interesting because, it being a
novel, its style is very different to that of Darwin's and Russell's
treatises. As in other novels, much of the structure is built up
around its characters, through a network of relationships that
change throughout the text. This is evidenced in the list of Table 1
by the prominence of pronouns and proper names, in addition to the
nouns that set the thematic focus of the book.   A small number of
words in the example listings have been assigned a large value of
information without being representative of the main subject of the
texts. That is likely due to fluctuations in the distributions of
those words, that affect the estimation of their information. We
have obtained listings like those of Table 1 for a variety of texts
in different languages, with highly consistent results. In all
cases, the words with the largest values of $\Delta I_w(s)$ were
among the most relevant to the respective subjects.  It is
remarkable that a simple measure based on Shannon mutual information
can readily identify the most important words in individual texts.

\section{Discussion}
The richness and diversity of linguistic structures pose a challenge
to the characterization of the complex information conveyed by
language using the quantitative techniques. Here we used a direct
application of information theory to quantify semantic information
in language sequences. Our main hypothesis was that the key
signatures of the semantic content of a text will be reflected in
long-range statistical patterns in the use of words, and thus should
be captured by basic information theoretic quantities. In
particular, we introduced a measure of relative information, $\Delta
I(s)$, that gauges a degree of order in written language based on
the statistical description of word frequencies in sections of the
text of a given size $s$.

For individual texts, the information $\Delta I(s)$ characterizes
the presence of an optimal scale $s$ at which the information
attains a maximum.  If a text is divided into sections of the size
of the optimal scale, the distribution of words in those parts will
be statistically the most diverse. For single texts with an overall
thematic unity, those divisions represent the typical spans in which
lines of argument and semantic structure develop.  An empirical
support for this interpretation comes from the inspection of
specific written works.  Books with the smallest optimal scales were
those made up of collections of disparate short text fragments. On
the contrary, the books with the largest optimal scales were long
treatises on well defined subjects. For texts such as Darwin's { \it
The Origin of Species}, Russell's {\it Analysis of the Mind}, and
Melville's {\it Moby Dick}, the optimal scales were in the range
between 1000 and 3000 words. This is roughly the length needed to
develop a line of argument in a literary or scholarly text. While it
can probably be natural to expect that complex messages in written
language build up on scales of a few thousands words, to the extent
of our knowledge, we proposed the first information-theoretic
quantification of those long range semantic structures.

The information $\Delta I(s)$ is built up from the contributions of
the information associated with individual words. The study of the
information per word, $\Delta I_w(s)$, revealed the presence of a
maximum at a scale that depended on the particular word. That
optimal scale is such that the distribution of the word over the
text partition differs the most from that of a word with the same
frequency but otherwise randomly distributed over the text. Using a
simple model of word distribution we showed that the optimal scale
for single words is related to the size of the typical domains, or
clusters, in which words tend to appear in written language.

Moreover, when words are ranked according to their information
contribution, the most informative ones are those clearly
representative of the message in the text. Remarkably, these most
informative words could be identified by using our
information-theoretical measure without any a priori knowledge of
the language or the text, apart from the identification of the
elementary information-carrying tokens –words themselves.

Probably, similar features can be found in other information
carriers in nature, where a particular kind of semantic information
leaves its mark on the long-range distribution of the basic symbols.
Much of the insight gained in our study could thus be extended to
the analysis of other biological information structures, like some
neural signals, the genetic code, and patterns of animal
communication and behavior.   Overall, our results suggest that
complex aspects of the information encoded in symbolic sequences are
susceptible of quantitative characterisation and analysis using the
rigorous principles of information theory.

\section{Acknowledgements}
We are grateful to In\'es Samengo for her critical reading of the
manuscript. This work was supported by the UK Medical Research
Council and the Royal Society (MAM), and CONICET and ANPCyT,
Argentina (DHZ).

\vspace*{-3pt}   
\bibliographystyle{ws-acs}

\vskip 2cm


{ \Large \bf Appendix A. Rate of word occurrence}
\vskip 0.75cm
In this section we give details about the calculation of the rate of
word occurrence shown in Fig.~ \ref{fig1} of the main text.

For a text of length $N$, we denote the positions of successive
words by an index $t=1,\ldots,N$.  The positions of a particular
word $w$, that appears $n$ times in the text, are then represented
by variables $t_i$, with $i=1,\ldots,n$. Thus, the density of
occurrences of word $w$ can be represented as
\begin{equation}
\nu_w(t)=\sum_{j=1} \delta(t-t_j)
\end{equation}
where $\delta(t)$ is the Dirac delta function \cite{arfken2005}. In
order to reveal the domain-like patterns in the distribution of
words, we define a rate of occurrence for each word as a convolution
of the density $\nu_w(t)$ with a bell-shaped kernel,
\begin{equation}
\rho_w(t)=\int_{-\infty}^{\infty}G(t-t_w',\sigma)\nu_w(t')dt'  ,
\end{equation}
where for simplicity of notation we assumed that the variable $t$ is continuous.
In particular, we used the following zero-mean Gaussian kernel:
\begin{equation}
G(t,\sigma)=\frac{e^{-\frac{t^2}{2\sigma^2}}}{\sqrt{2\pi} \sigma} ,
\end{equation}
where the parameter $\sigma$ controls the width. It can be shown
that this type of kernel introduces correlations in the rate
$\rho_w(t)$ over spans of order of $\sigma$. To obtain the data for
Fig. \ref{fig1} we used $\sigma=50$ words.  Therefore, all the
structure show in the figure can be attributed to genuine
correlations in the distributions of words.

\vskip 2cm
{ \Large \bf Appendix B. Derivation of $\Delta I(s)$ } 
\vskip 0.75cm
In this section we supply additional details on the derivation of
the information measure $\Delta I(s)$, given by Eq.\ref{DI}.

Let us consider a given word $w$, which appears $n_j$ times in part
$j$, with $j=1,\ldots, P$. We can then define the conditional
probability of finding word $w$ in part $j$, as $p(w|j)=n_j/N_j$.
If $K$ is the size of the lexicon, the normalization condition for
the above probability is $\sum_{w=1}^K p(w|j)=1$. If we now call
$p(j)=N_j/N$ the {\it a priori} probability that the  word $w$
appears in part $j$, then $\sum_{j=1}^P p(w|j)p(j)=p(w)$, where
$p(w)=n/N$ stands for the overall probability of occurrence of a
word in the whole text.

The probability $p(w|j)$ tells us how likely is to find word $w$
given that we are looking into part $j$. Here we are interested in
assessing how well a particular word tags a section, or part, in a
text. Such information is given by the inverted probability
$p(j|w)$, which tells how likely is that we are looking into part
$j$ given that we saw an instance of word $w$ in the text. This
probability is easily found by means of Bayes's rule
\cite{feller_book}, as follows:
\begin{equation}
\label{bayes} p(j|w)=\frac{p(w|j)p(j)}{\sum_{i=1}^P p(w|i)p(i)}  .
\end{equation}
Then, by writing explicitly the probabilities in the right-hand side
of Eq. (\ref{bayes}), we find that $p(j|w)=n_j/n$.

With the previous definitions we can write Shannon mutual
information between the text sections and the words
\cite{cover2006}:
\begin{equation} M(J,W)=\sum_{w=1}^K
p(w)\sum_{j=1}^P p(j|w) \log_2 \left(\frac{p(j|w)}{p(j)}\right)  .
\end{equation}

However, as was discussed in the main text, we are interested in a
measure that quantifies the specific linguistic information in the
long-range distribution of words. We therefore need to subtract the
residual information contributed by fluctuations in word frequencies
over the different parts. These fluctuations are particularly
important for words that occur a number of times$n << N$
\cite{montemurro_zanette2002,schurmann1996}.

Let us call $\hat{M}(J,W)$ the mutual information computed on one
particular random realization of the text obtained by shuffling all
of the words' positions. In order to obtain a representative
quantity that does not depend on one particular realization of the
random shuffling, we take an average over all the possible
realizations of the random text, represented as $\langle
\hat{M}(J,W) \rangle$.  We can now define the information in the
distribution of words simply as the difference in the value of the
mutual information calculated on the real and the average over the
random texts, $\Delta(s)=M(J,W)-\langle \hat{M}(J,W) \rangle$, where
we made explicit the dependency on the scale parameter $s$. After
expanding and regrouping terms we can write the relative information
$\Delta I(s)$ as follows:
\begin{equation}
\label{DIs} \Delta I(s)=\sum_{w=1}^K p(w)\left[\langle
\hat{H}(J|w)\rangle-H(J|w)\right]  ,
\end{equation}

The first of the entropies appearing in Eq. (\ref{DIs}) corresponds
to the one computed on the random text and averaged over all the
realizations of the random permutation of words, and is defined as
follows:
\begin{equation}
\label{entropyhat} \langle \hat{H}(J|w)\rangle=-\sum_{j=1}^P \langle
\hat{p}(j|w)\log_2 \hat{p}(j|w)\rangle  ,
\end{equation}
where we denote by $\hat{p}(j|w)$ the probabilities computed on the
random text. The second entropy is estimated directly on the real
text, and its definition is the following:
\begin{equation}
H(J|w)=-\sum_{j=1}^P p(j|w)\log_2 p(j|w)  .
\end{equation}

In our approach, the amount of information contributed by single
words to the total information is proportional to the difference in
entropies for the word in the random version of text and the one on
the real text. The proportionality factor is given by the a weight
equal to the overall probability of a word in the text, $p(w)$.
\vskip 2cm
{ \Large \bf Appendix C. Analytic computation of the entropy of the random text} 
\vskip 0.75cm

Here we compute an analytic expression for the entropy $\langle
\hat{H}(J|w)\rangle$, appearing in Eq. (\ref{DIs}). As was discussed
above this entropy is computed on a stochastic version of the text
obtained by randomly shuffling all the words' positions. The
brackets $\langle \ldots \rangle$ denote an average over all
possible randomly shuffled texts.

Again, let us suppose that the random text has a length of $N$
words, and that it is divided in $P$ parts of equal length. Using
Eq. (\ref{entropyhat}), for a word that appears $m_j$ times in part
$j$ with a frequency $n$ over the whole text, this entropy takes the
following form:
\begin{equation}
\hat{H}(J|w)=-\sum_{j=1}^P \frac{m_j}{n}\log_2 \frac{m_j}{n}  .
\end{equation}

The average of the entropy $\hat{H}(J|w)$ over all possible
realizations of the random text is computed as follows:
\begin{equation}
\label{hatav} \langle \hat{H}(J|w)\rangle=-\sum_{\begin{array}{cc}
m_1+\ldots +m_P=n &
 \\ \mbox{with}\,    m_j\le N/P , j=1,\ldots,P &  \end{array}}
 p(m_1,\ldots ,m_P)\sum_{j=1}^P \frac{m_j}{n}\log_2 \frac{m_j}{n} ,
\end{equation}
where  $p(m_1,\ldots ,m_P)$ is the probability of finding $m_j$
words in part $j$, with $j=1,\ldots,P$. We notice that in Eq.
(\ref{hatav}), for each of the $P$ terms in the second sum, the
first sum can be taken over all indices except one. This leads to
the following simpler form of the average entropy:
\begin{equation}
\langle \hat{H}(J|w)\rangle=-P \sum^{\mbox{min}\{n,N/P\}}_{m=1}p(m)
\frac{m}{n}\log_2 \frac{m}{n} .
\end{equation}

Finally, the marginal probability $p(m)$ can now be easily computed.
It is given by the probability of finding $m$ instances of word $w$
in one part, together with $N/P-m$ words different from $w$, and
reads
\begin{equation}
p(m)=\frac{ {n \choose m}{N-n \choose N/P-m}}{{N \choose N/P}}.
\end{equation}

\end{document}